\documentclass[epj]{webofc}
\usepackage[varg]{txfonts}   
\wocname{EPJ Web of Conferences}
\woctitle{CONF12}
\newcommand{\cO}{\mathcal{O}}

\newcommand{\be}{\begin{equation}}
\newcommand{\ee}{\end{equation}}

\begin{document}
\selectlanguage{english}
\title{Precision physics with QCD}
%
%

\author{Antonio Pich\inst{1}\fnsep\thanks{\email{Antonio.Pich@ific.uv.es}}}

\institute{Departament de F\'{\i}sica Te\`orica, IFIC, Universitat de Val\`encia -- CSIC\\ Edifici d'Instituts de Paterna, Apt. Correus 22085, E-46071, Val\`encia, Spain}

\abstract{%
The four-loop determination of the strong coupling from fully inclusive observables is reviewed. Special attention is given to the low-energy measurement extracted from the hadronic $\tau$ decay width. A recent exhaustive analysis of the ALEPH data, exploring several complementary methodologies with very different sensitivities to inverse power corrections and duality violations, confirms the strong suppression of non-perturbative contributions to $R_\tau$. 
It gives the value $\alpha_s(m_\tau^2)= 0.328 \pm 0.013$, which implies 
$\alpha_s(M_Z^2)= 0.1197 \pm 0.0015$. The excellent agreement with the direct measurement at the $Z$ peak, $\alpha_s(M_Z^2)= 0.1196 \pm 0.0030$, provides a beautiful test of asymptotic freedom. Together with the most recent lattice average from FLAG and the NNLO determinations from $e^+e^-$, PDFs and collider data quoted by the PDG, these two inclusive determinations imply a world average value  
$\alpha_s(M_Z^2)= 0.1180 \pm 0.0010$.  
}
\maketitle
\section{Introduction}
\label{sec:intro}

All strong interaction phenomena should be described in terms of the strong coupling 
$\alpha_s$, the single free parameter of Quantum Chromodynamics (QCD).
The overwhelming consistency of the many determinations of $\alpha_s$, performed in different processes and at different mass scales provides a beautiful verification of QCD. 
A good understanding of the uncertainties associated with the different measurements is needed in order to appreciate the significance of this test, which must be then restricted to observables where perturbative techniques are reliable and enough terms in the perturbative expansion are available. The PDG \cite{Olive:2016xmw} requires a NNLO (or higher) theoretical accuracy. In addition, small non-perturbative corrections are always present, specially at low energies, and one should also worry about the expected asymptotic behaviour of the perturbative series.

The most reliable determinations of $\alpha_s$ have been compiled in 
Refs.~\cite{BDS:16,d'Enterria:2015toz,Pich:2013sqa,Bethke:2012jm,Bethke:2011tr}. 
I will focus the discussion on the very precise inclusive observables $R_Z$ and $R_\tau$, which are already known to four loops, {\it i.e.}, to N${}^3$LO, and
will update the PDG information with the most recent developments, not yet included in the official averages.

\section{Running coupling and effective QCD theories}
\label{sec:running}

The QCD coupling obeys the renormalization group equation
\begin{equation}
\mu\,\frac{d\alpha_s(\mu^2)}{d\mu}\; =\; \alpha_s(\mu^2)\,\beta(\alpha_s)\, ,
\qquad\qquad\quad
\beta(\alpha_s)\; =\;\sum_{n=1}\, \beta_n\, a_s^n\, ,
\qquad\qquad\quad
a_s\; =\;\frac{\alpha_s}{\pi}\, .
\end{equation}
The fifth-order coefficient of the $\beta$ function has been recently computed in Ref.~\cite{Baikov:2016tgj} (see also Ref.~\cite{Luthe:2016ima}), which provides a quite precise perturbative control of the scale dependence of $\alpha_s$. In the $\overline{\mathrm{MS}}$ scheme ($\beta_1$ and $\beta_2$ are scheme independent), the known coefficients are  \cite{Baikov:2016tgj,vanRitbergen:1997va,Czakon:2004bu}:
%
\begin{eqnarray}
&&\mbox{}\hskip -1.1cm \beta_1\, =\,\frac{1}{3}\, n_f - \frac{11}{2}\, ,
\qquad\quad
\beta_2\, =\, -{51\over 4} + {19\over 12} \, n_f \, ,
\qquad\quad
\beta_3 \, = \, {1\over 64}\left[ -2857 + {5033\over 9} \, n_f
- {325\over 27} \, n_f^2 \right]\,  ,
\nonumber\\[3pt]
&&\mbox{}\hskip -1.1cm \beta_4\, =\,
\frac{-1}{128}\, \left[
\frac{149753}{6}+3564 \,\zeta_3 
-\left( \frac{1078361}{162}+\frac{6508}{27}\,\zeta_3 \right)\, n_f
+ \left( \frac{50065}{162}+\frac{6472}{81}\,\zeta_3 \right)\, n_f^2
+\frac{1093}{729}\, n_f^3 \right]\, ,
\nonumber\\[3pt]
&&\mbox{}\hskip -1.1cm \beta_5\, =\,
-\frac{1}{512}\,\Biggl\{
\frac{8157455}{16} +\frac{621885}{2} \,\zeta_{3} -\frac{88209}{2} \,\zeta_{4}
-288090 \,\zeta_{5}
\nonumber\\
&& \hskip .5cm \mbox{} + n_f\:\left[
-\frac{336460813}{1944} 
-\frac{4811164}{81}  \, \zeta_{3}
+\frac{33935}{6}  \, \zeta_{4}
+\frac{1358995}{27}  \, \zeta_{5}
\right]
\nonumber\\
&& \hskip .5cm \mbox{} + n_f^2\:\left[
\frac{25960913}{1944} 
+\frac{698531}{81}  \, \zeta_{3}
-\frac{10526}{9}  \, \zeta_{4}
-\frac{381760}{81}  \, \zeta_{5}
\right]
\nonumber\\
&& \hskip .5cm \mbox{} + n_f^3\:\left[
-\frac{630559}{5832} 
-\frac{48722}{243}  \, \zeta_{3}
+\frac{1618}{27}  \, \zeta_{4}
+\frac{460}{9}  \, \zeta_{5} \right]
\, +\, n_f^4\:\left[ \frac{1205}{2916} -\frac{152}{81}  \, \zeta_{3} \right]\,
\Biggr\}\, .
\end{eqnarray}
The very modest growth of $\beta_n$ with the perturbative order gives rise to a surprisingly smooth power expansion. For $n_f=5$, for instance,
$\beta(\alpha_s) =  \beta_1 a_s\,\left( 1 + 1.26\, a_s + 1.47\, a_s^2 + 9.83\, a_s^3 + 7.88\, a_s^4\right)$.

\begin{figure}[t]
\centering
\includegraphics[width=\textwidth,clip]{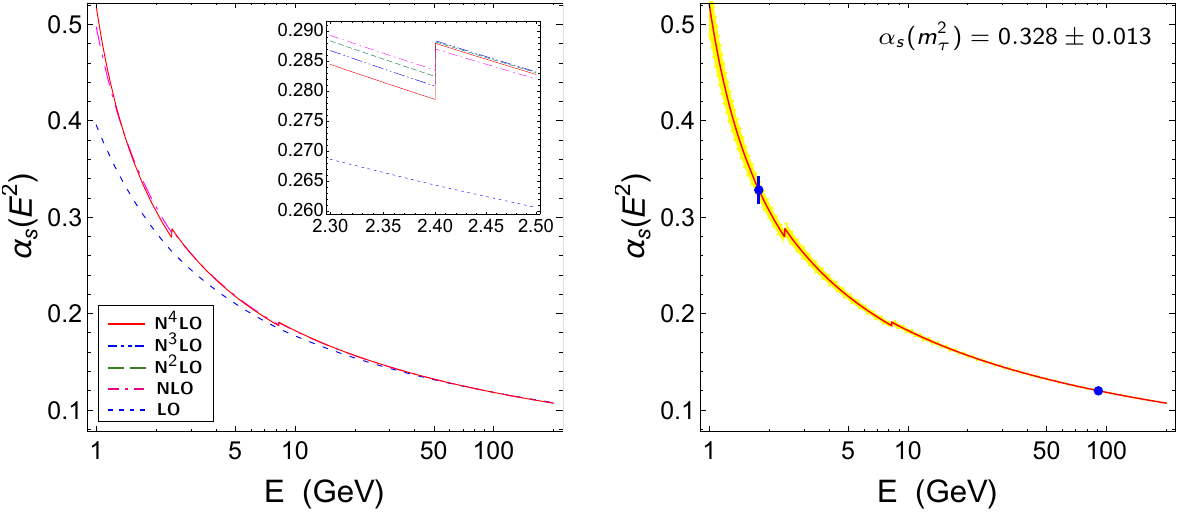}
\caption{Scale dependence of $\alpha_s$ at different perturbative orders (left).
The right plot compares the 5-loop evolution of $\alpha_s(m_\tau^2)$, determined from hadronic $\tau$ decays, with the measurement of $\alpha_s(M_Z^2)$ from $\Gamma_Z$.} 
\label{fig:running}
\end{figure}

The scale dependence of $\alpha_s$ over a wide range of energies, at different levels of approximation, is shown in figure~\ref{fig:running}.
The 5-loop precision in the $\beta$ function implies a resummation of N${}^4$LO logarithmic contributions to the running of $\alpha_s$, {\it i.e.}, corrections of the form $\Delta\alpha_s(Q^2)\sim \alpha_s(\mu^2)^{n+5} \log^n{(Q^2/\mu^2)}$. 
Owing to the fast convergence of the $\beta$ function, the NLO resummation gives already an excellent approximation to the running coupling.  The achieved accuracy is quite impressive; the four and five loop corrections are so small that it is difficult to appreciate them in the figure. 

The small discontinuities in the plotted curves reflect the crossing of the charm and bottom thresholds where one needs to properly match the different QCD${}_{n_f}$ effective theories. Since the $\beta_n$ coefficients are functions of $n_f$, the strong coupling depends 
on the considered number of ``active'' quark flavours. When a quark is heavy enough to decouple, it is convenient to remove it from the Lagrangian and work with an effective QCD theory which has one quark less and a different value of $\alpha_s$.
The matching conditions relating the effective QCD theories with $n_f$ and $n_f-1$ flavours are known to four loops \cite{Schroder:2005hy,Chetyrkin:2005ia}.

\section{Inclusive observables}
\label{sec:inclusive}

Inclusive observables, such as $\sigma(e^+e^-\to\mathrm{hadrons})$ at high-enough energies, $\Gamma(Z\to\mathrm{hadrons})$ or $\Gamma(W\to\mathrm{hadrons})$, can be accurately predicted  with perturbative methods. Since the final hadrons are produced through the vector
$\, V^{\mu}_{ij} = \bar{\psi}_j \gamma^{\mu} \psi_i \, $
and axial-vector
$\, A^{\mu}_{ij} = \bar{\psi}_j \gamma^{\mu} \gamma_5 \psi_i \,$
colour-singlet quark currents ($i,j=u,d,s\ldots$),
the QCD dynamics is governed by the two-point correlation functions
\begin{equation}\label{eq:pi_v2}
\Pi^{\mu \nu}_{ij,J}(q)\; \equiv\;
 i \int d^4x \;\, \mathrm{e}^{iqx}\,
\langle 0|T(J^{\mu}_{ij}(x)\, J^{\nu}_{ij}(0)^\dagger)|0\rangle
\; =\;
\left( -g^{\mu\nu} q^2 + q^{\mu} q^{\nu}\right) \, \Pi_{ij,J}^{(0+1)}(q^2)
 +   g^{\mu\nu} q^2\, \Pi_{ij,J}^{(0)}(q^2) \, ,
\end{equation}
where $J=V,A$ and the superscript $L=0,1$ denotes the angular momentum in the hadronic rest frame. The correlators $\Pi_{ij,J}^{(L)}(q^2)$ are analytic functions of $q^2$, in the complex $q^2$ plane, except along the (physical) positive real axis where their imaginary parts have discontinuities which correspond to the measurable hadronic spectral distributions with the given quantum numbers.

For massless quarks, $s\,  \Pi_{ij,J}^{(0)}(s)= \mathrm{constant}$ (there is a non-perturbative Goldstone-pole contribution to $\Pi_{ij,A}^{(0)}$ at $s=0$, which cancels in $\Pi_{ij,A}^{(0+1)}$).  
When $i\not= j$, the two quark currents must necessarily be connected through a quark loop
(non-singlet topology), which gives identical contributions to the vector and axial massless correlators:\
$\Pi(s)\equiv \Pi_{i\not=j,V}^{(0+1)}(s)=\Pi_{i\not=j,A}^{(0+1)}(s)$.
They are conveniently parametrized through the Euclidean Adler function ($Q^2=-q^2$ and $N_C=3$ is the number of quark colours)
\begin{equation}\label{eq:Adler}
D(Q^2)\; \equiv\; -Q^2\frac{d}{dQ^2}\Pi(Q^2)\; =\; \frac{N_C}{12\pi^2}\;\left\{
1 + \sum_{n=1}\; K_n
\left( {\alpha_s(Q^2)\over \pi}\right)^n\right\}\, ,
\end{equation}
which is known to $\mathcal{O}(\alpha_s^4)$ \cite{Baikov:2008jh,Gorishnii:1990vf,Surguladze:1990tg}:
\begin{eqnarray}
\lefteqn{\hskip -.5cm K_1 \, =\, 1\, ,
\qquad
K_2\, =\, 1.98571 -0.115295\; n_f\, ,
\qquad
K_3\, =\, 18.2427 - 4.21585\; n_f + 0.0862069\; n_f^2\, ,}&& 
\nonumber\\
\lefteqn{\hskip -.5cm K_4 \, =\, 135.792 - 34.4402\; n_f + 1.87525\; n_f^2 - 0.0100928\; n_f^3\, .}&&
\end{eqnarray}
There are additional singlet contributions to the neutral-current correlators ($i=j$), with
each current coupling to a different quark loop. Since gluons have $J^{PC}=1^{--}$ and colour, these topologies start to contribute at $\mathcal{O}(\alpha_s^3)$ and $\mathcal{O}(\alpha_s^2)$, respectively, for the vector and axial-vector currents:
\begin{eqnarray}
\Delta^\mathrm{S} D_V(Q^2)\; =\; \frac{N_C}{12\pi^2}\;\sum_{n=3}\; d^{V}_n
\left( {\alpha_s(Q^2)\over \pi}\right)^n ,
\qquad\quad
\Delta^\mathrm{S} D_A(Q^2)\; =\; \frac{N_C}{12\pi^2}\;\sum_{n=2}\; d^{A}_n
\left( {\alpha_s(Q^2)\over \pi}\right)^n .
\end{eqnarray}
The vector-current coefficients are
$d_3^V = -0.41318$ and $d_4^V = -5.94225 + 0.191628\; n_f$ \cite{Baikov:2012er}.

\subsection{$\boldsymbol{\sigma(e^+e^-\to\mathbf{hadrons})}$}
\label{subsec:Ree}

The ratio of the electromagnetic $e^+e^-\to\mathrm{hadrons}$ and $e^+e^-\to\mu^+\mu^-$
cross sections is given by
\begin{eqnarray}\label{eq:R_ee}
R_{e^+e^-}(s)& \equiv&
\frac{\sigma(e^+e^-\to \mathrm{hadrons})}{\sigma(e^+e^-\to\mu^+\mu^-)}\;
=\; 12 \pi \;\left\{ \sum_f Q_f^2\;\, \mathrm{Im} \Pi(s)
+ \left(\sum_f Q_f\right)^{\! 2}\; \mathrm{Im}\, \Delta^\mathrm{S}\Pi_V(s)\right\}
\nonumber\\
& =& \sum_{f} Q_f^2\; N_C \; \left\{ 1 +
\sum_{n\geq 1} F_n \left({\alpha_s(s)\over\pi}\right)^{\! n} \right\}
\; + \; \mathcal{O}\left(\frac{m_q^2}{s},\frac{\Lambda^4}{s^2}\right)\, .
\end{eqnarray}
The sum over quark electric charges of different signs strongly suppresses the singlet contribution, which has been included as a small correction to the coefficients $F_{n\ge 3}$. For $n_f=5$ flavours, one gets\ $F_1 = 1$, $F_2 = 1.4092$,
$F_3 = -12.805$ and $F_4 = -80.434$ \cite{Baikov:2012er}.

The perturbative series in Eq.~(\ref{eq:R_ee}) is actually an expansion in powers of $\alpha_s(\mu^2)$ with coefficients containing a polynomial dependence on $\log{(s/\mu^2)}$. These logarithms are resummed into the running coupling by taking $\mu^2=s$. Although the physical ratio $R_{e^+e^-}(s)$ is independent of the renormalization scale $\mu$, the truncated series contains a residual $\mu$-dependence of $\mathcal{O}(\alpha_s^{N+1})$, where $N=4$ is the last included term, which must be taken into account in the theoretical uncertainty. Since non-perturbative corrections are suppressed by $\Lambda^4/s^2$ (the gauge-invariant operators contributing to the current correlators have dimensions $D\ge 4$), at high energies one can perform a N${}^3$LO determination of $\alpha_s(s)$. Unfortunately, the experimental uncertainties are large.

\subsection{$\boldsymbol{\Gamma(Z\to\mathrm{\bf hadrons})}$}
\label{subsec:Zwidth}

The electroweak neutral current
$J^\mu_Z = \sum_f (v_f V_{ff}^\mu + a_f A_{ff}^\mu)$ contains vector and axial-vector components, weighted with the corresponding $Z$ couplings. The singlet axial contributions of the two members of a weak isospin doublet cancel each other for equal quark masses because $a_f = 2 I_f$; however, the large value of the top mass generates very important
singlet axial corrections which start at $\mathcal{O}(\alpha_s^2)$. The ratio of the hadronic and electronic widths of the $Z$ boson involves the QCD series ($m_b=0$, $m_t\not=0$)
\begin{equation}\label{eq:RZ}
R_Z\; \equiv\; \frac{\Gamma(Z\to\mathrm{hadrons})}{\Gamma(Z\to e^+e^-)}\; =\; R_Z^{\mathrm{EW}} \; N_C\;\left\{
1 + \sum_{n=1}\; \tilde F_n
\left( {\alpha_s(M_Z^2)\over \pi}\right)^n\right\}
\, ,
\end{equation}
with $\tilde F_1 = 1$, $\tilde F_2 = 0.76264$,
$\tilde F_3 = -15.490$ and $\tilde F_4 = -68.241$ \cite{Baikov:2012er}.
Taking properly into account the electroweak corrections and QCD contributions suppressed by powers of $m_b^2/M_Z^2$ \cite{Chetyrkin:1994js,Chetyrkin:2000zk}, the ratio $R_Z$ is included in the global fit to electroweak precision data. This results in a quite accurate value of $\alpha_s(M_Z^2)$ \cite{Baak:2014ora}:
\begin{equation}\label{eq:alpha_Z}
\alpha_s^{(n_f=5)}(M_Z^2)\; \equiv\; \alpha_s(M_Z^2)\; =\; 0.1196\pm 0.0030\, .
\end{equation}
This determination assumes the validity of the electroweak Standard Model. 

\section{Hadronic decay width of the $\boldsymbol{\tau}$ lepton}

\begin{figure}[t]
\centerline{
\includegraphics[width=0.36\textwidth]{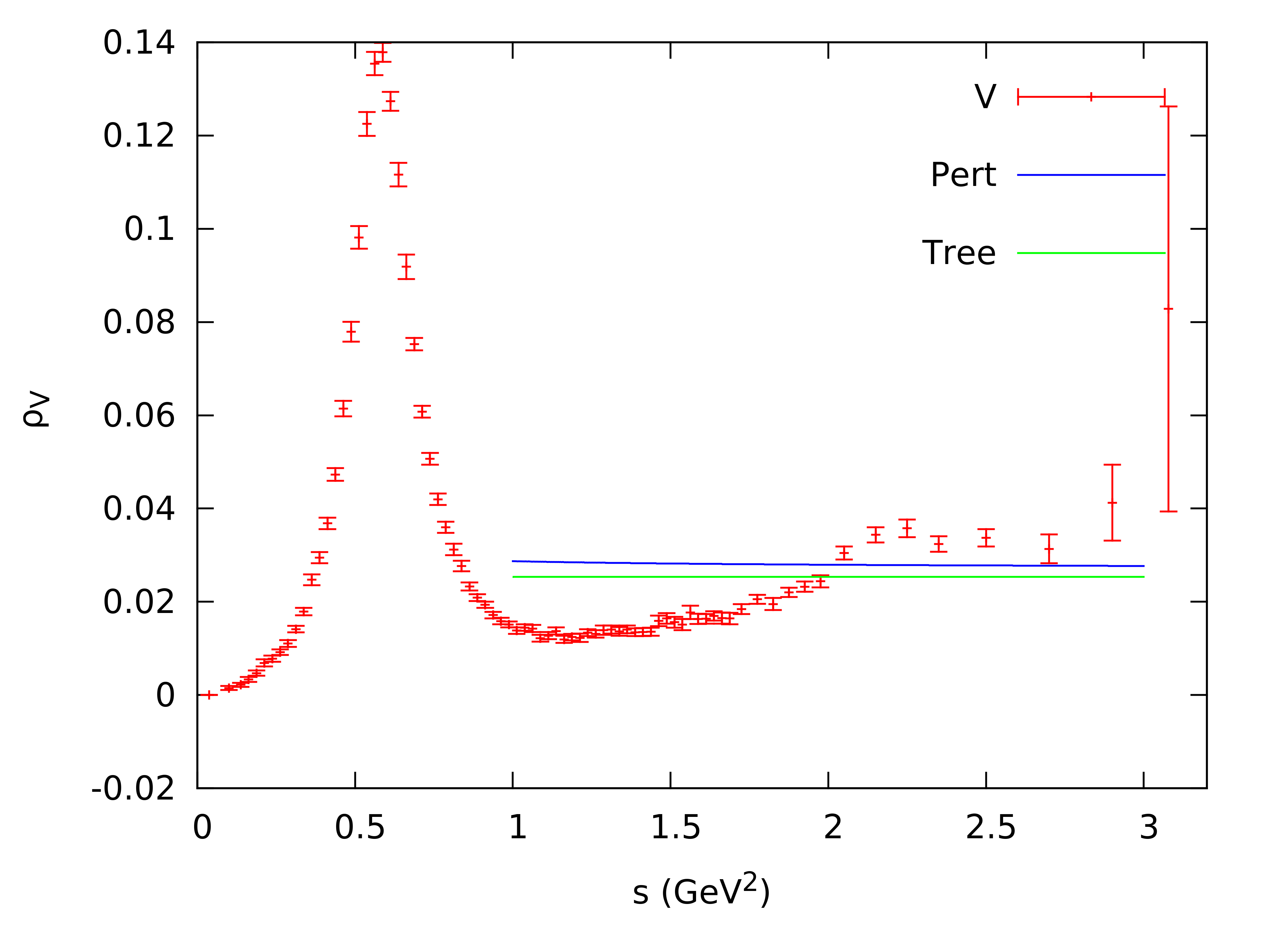}\hskip -.15cm
\includegraphics[width=0.36\textwidth]{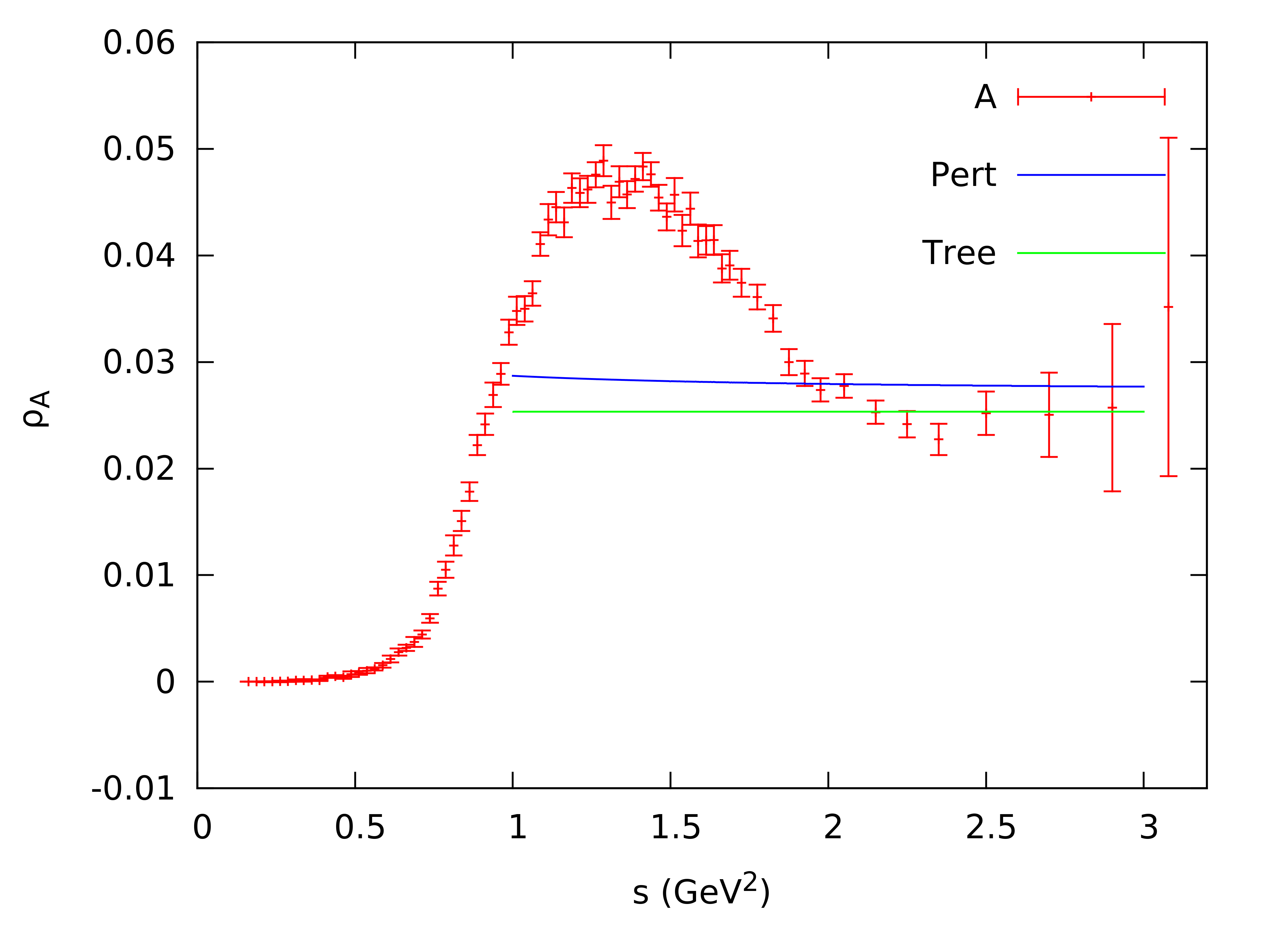}\hskip -.15cm
\includegraphics[width=0.36\textwidth]{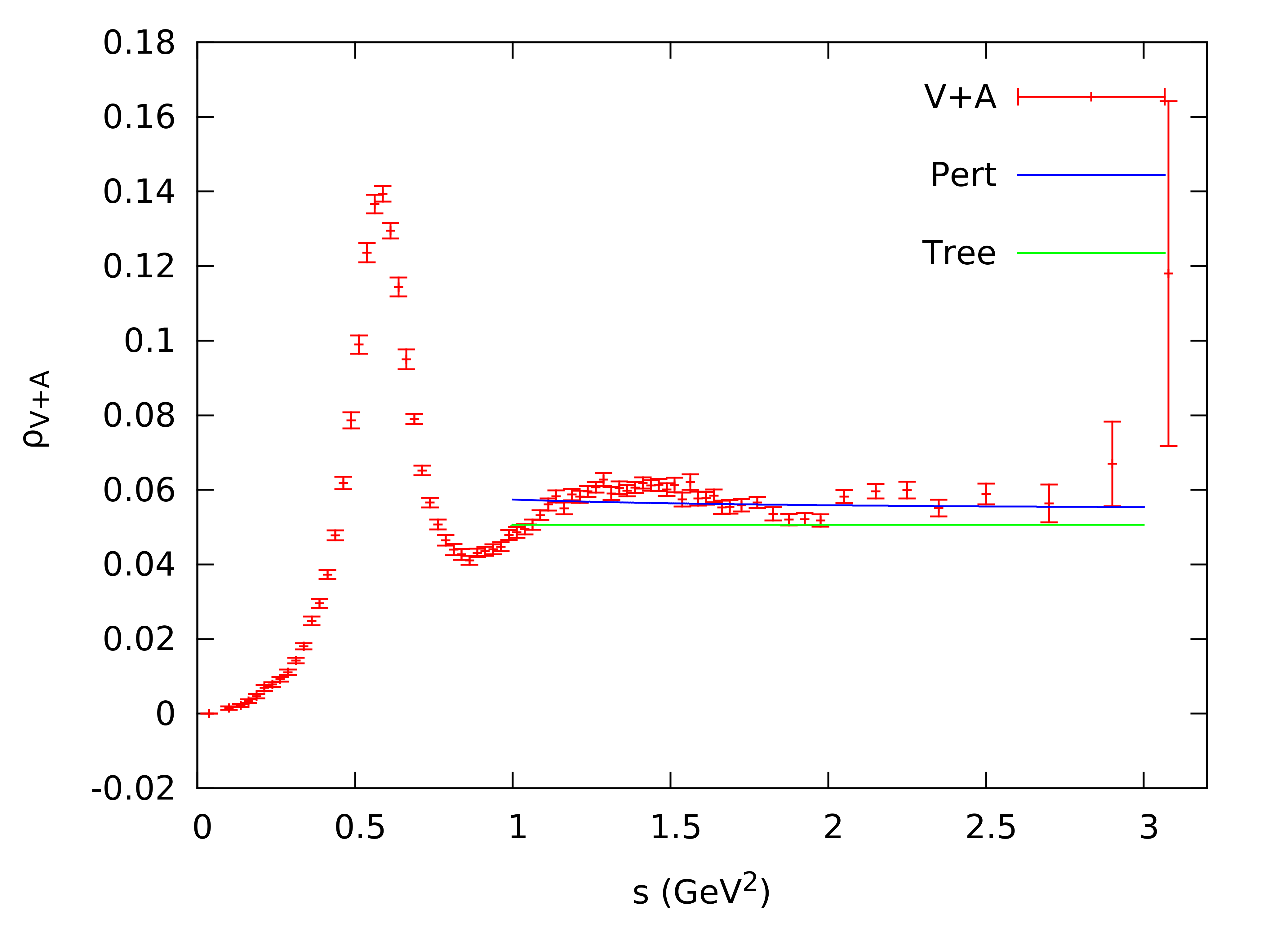}
}
\caption{Spectral functions for the $V$, $A$ and $V+A$ channels, determined from ALEPH $\tau$ data \cite{Davier:2013sfa}.}
\label{fig:SpectralFunction}
\end{figure}

The hadronic $W^\pm$ decay width does not provide yet a competitive determination of $\alpha_s$. A much better alternative \cite{Narison:1988ni,Braaten:1988hc,Braaten:1991qm} is the hadronic $\tau$ decay, which proceeds through a virtual $W^\pm$ boson. The QCD correlation function of two left-handed charged currents receives only non-singlet contributions. Restricting the analysis to the dominant Cabibbo-allowed decay width,
\begin{eqnarray}\label{eq:R_tau}\hskip -.5cm
R_{\tau,V+A} &\!\!\equiv &\!\!\frac{ \Gamma [\tau^- \to \nu_\tau +\mathrm{hadrons}\, (S=0)]}{ \Gamma [\tau^- \to \nu_\tau e^- {\bar \nu}_e]}
\\ \hskip -.5cm\mbox{}
&\!\! = &\!\! 12 \pi\, |V_{ud}|^2\, S_{\mathrm{EW}} \int^{m_\tau^2}_0 {ds \over m_\tau^2 } \,
 \left(1-{s \over m_\tau^2}\right)^2
\biggl[ \left(1 + 2 {s \over m_\tau^2}\right)
 \mbox{\rm Im} \Pi^{(0+1)}_{ud,V+A}(s)
 - 2 {s \over m_\tau^2}\, \mbox{\rm Im} \Pi^{(0)}_{ud,V+A}(s) \biggr]\,  ,
\nonumber\end{eqnarray}
where $S_{\mathrm{EW}}=1.0201\pm 0.0003$ incorporates the
electroweak radiative corrections \cite{Marciano:1988vm,Braaten:1990ef,Erler:2002mv}.
The measured invariant-mass distribution of the final hadrons determines the
spectral functions $\rho_J(s) \equiv \frac{1}{\pi}\,\mathrm{Im} \Pi_{ud,J}^{(0+1)}(s)$, shown in figure~\ref{fig:SpectralFunction} (the only relevant contribution to the $s\,\mathrm{Im} \Pi^{0}_{ud,V+A}(s)$ term is the $\pi^-$ final state at $s=m_\pi^2$).

Using the analyticity properties of the $\Pi^{(L)}_{ij,J}(s)$ correlators,
the experimental spectral distribution can be related with theoretical QCD predictions through moments of the type \cite{Braaten:1991qm,LeDiberder:1992zhd}
\be\label{aomega}
A^{\omega}_J(s_{0})\;\equiv\; \int^{s_{0}}_{s_{\mathrm{th}}} \frac{ds}{s_{0}}\;\omega(s)\; \mathrm{Im} \Pi_{ud,J}^{(0+1)}(s)\; =\; \frac{i}{2}\;\oint_{|s|=s_{0}}
\frac{ds}{s_{0}}\;\omega(s)\, \Pi_{ud,J}^{(0+1)}(s)\, ,
\ee
where $s_{\mathrm{th}}$ is the hadronic mass-squared threshold, $\omega(s)$ is any weight function analytic in $|s|\le s_0$, and the complex integral in the right-hand side (rhs) runs counter-clockwise around the circle $|s|=s_{0}$.
For large-enough values of $s_0$, the operator product expansion (OPE)
\be\label{eq:ope}
\Pi_{ud,J}^{(0+1)}(s)^{\mathrm{OPE}}\; =\; \sum_{D}\frac{1}{(-s)^{D/2}}\sum_{\mathrm{dim} \, \mathcal{O}=D} C_{D, J}(-s,\mu)\;\langle\mathcal{O}(\mu)\rangle
\;\equiv\; \sum_{D}\;\dfrac{\mathcal{O}_{D,\, J}}{(-s)^{D/2}}\, ,
\ee
can be used to predict the rhs integral as an expansion in inverse powers of $s_0$ (the $D=0$ term contains the perturbative contribution), while the lhs is directly determined by the experimental data.

The ratio $R_{\tau,V+A}$ in Eq.~(\ref{eq:R_tau}) corresponds to the particular weight 
$\omega(x) = (1-x^2)(1+2x)=1-3x^2+2x^3$, with $x\equiv s/s_0$ and $s_0=m_\tau^2$.
Thus, owing to Cauchy's theorem, the contour integral is only sensitive to OPE corrections with $D=6$ and 8, which are strongly suppressed by the corresponding powers of the $\tau$ mass (there is in addition a further suppression of the $D=6$ term because the vector and axial-vector contributions have opposite signs, cancelling to a large extent).
Moreover, $\omega(s)$ contains a double zero at $s=s_0$ which heavily suppresses the contribution to the integral from the region near the real axis, where the OPE is not valid.  This makes  $R_{\tau,V+A}$ a very clean observable to measure $\alpha_s$. It is very sensitive to the strong coupling because $\alpha_s(m_\tau^2)$ is sizeable, and non-perturbative effects are smaller than the perturbative uncertainties.

The availability of good experimental data makes possible to determine the small non-perturbative corrections from the data themselves, using weights with different powers of $s$ which are sensitive to the corresponding power corrections in the OPE \cite{LeDiberder:1992zhd}. The dominant uncertainty in the $\alpha_s(m_\tau^2)$ determination comes from the perturbative error associated with the unknown higher-order corrections to the Adler series in Eq.~(\ref{eq:Adler}). For a given value of $\alpha_s$, the so-called contour-improved perturbation theory (CIPT) \cite{LeDiberder:1992jjr,Pivovarov:1991rh}, which resumms large corrections arising from the long running along the circle $s=s_0$, results in a smaller perturbative contribution than the truncated fixed-order perturbation theory (FOPT) approximation \cite{Braaten:1991qm}. Therefore, CIPT leads to a larger fitted value of $\alpha_s(m_\tau^2)$ than~FOPT.

\subsection{Numerical analysis}

A detailed reanalysis of the $\alpha_s(m_\tau^2)$ determination from $\tau$ decay has been recently performed \cite{Pich:2016bdg}, including many consistency checks to assess the potential size of non-perturbative effects. All strategies adopted in previous works have been investigated, studying the stability of the results and trying to uncover any potential hidden weaknesses, and several complementary approaches have been considered.
Once their uncertainties are properly estimated, all adopted methodologies result in very consistent values of $\alpha_s(m_\tau^2)$. Table~\ref{tab:summary} summarizes the
most reliable determinations.

All analyses have been done both in CIPT and FOPT. Within a given approach the perturbative errors have been estimated varying the renormalization scale in the interval $\mu^2/s_0 \in [0.5\, ,\, 2]$, and taking $K_{5}=275 \pm 400$ as an educated guess of the maximal range of variation of the unknown fifth-order contribution \cite{Pich:2011bb}. These two sources of theoretical uncertainty have been combined in quadrature, together with the experimental errors. The different values quoted in the table include, as an additional uncertainty, the variations of the results under various modifications of the fit procedures. The systematic difference between the values obtained with the CIPT and FOPT prescriptions appears clearly manifested in the table. The CIPT and FOPT results have been finally averaged, but adding in quadrature half their difference to the smallest of the CIPT and FOPT errors.

\begin{table}[t]
\centering
\caption{Determinations of $\alpha_{s}^{(n_f=3)}(m_{\tau}^{2})$ from $\tau$ decay data, with different methods, in the $V+A$ channel \protect\cite{Pich:2016bdg}.}
\label{tab:summary}
\begin{tabular}{|c|c|c|c|}
\hline &\multicolumn{3}{c|}{}\\[-10pt]
Method  & \multicolumn{3}{c|}{$\alpha_{s}(m_{\tau}^{2})$}
\\[1.5pt] \cline{2-4}
& \raisebox{-2pt}{CIPT} & \raisebox{-2pt}{FOPT} & \raisebox{-2pt}{Average}
\\[1.2pt] \hline &&&\\[-10pt]
ALEPH moments & $0.339 \,{}^{+\, 0.019}_{-\, 0.017}$ &
$0.319 \,{}^{+\, 0.017}_{-\, 0.015}$ & $0.329 \,{}^{+\, 0.020}_{-\, 0.018}$
\\[3pt]
Modified ALEPH moments  & $0.338 \,{}^{+\, 0.014}_{-\, 0.012}$ &
$0.319 \,{}^{+\, 0.013}_{-\, 0.010}$ & $0.329 \,{}^{+\, 0.016}_{-\, 0.014}$
\\[3pt]
$A^{(2,m)}$ moments  & $0.336 \,{}^{+\, 0.018}_{-\, 0.016}$ &
$0.317 \,{}^{+\, 0.015}_{-\, 0.013}$ & $0.326 \,{}^{+\, 0.018}_{-\, 0.016}$
\\[3pt]
$s_0$ dependence  & $0.335 \pm 0.014$ &
$0.323 \pm 0.012$ & $0.329 \pm 0.013$
\\[3pt]
Borel transform  & $0.328 \, {}^{+\, 0.014}_{-\, 0.013}$ &
$0.318 \, {}^{+\, 0.015}_{-\, 0.012}$ & $0.323 \, {}^{+\, 0.015}_{-\, 0.013}$
\\[2pt] \hline
\end{tabular}
\end{table}

The first determination in table~\ref{tab:summary} follows the method adopted in the ALEPH analysis of Ref.~\cite{Davier:2013sfa}, taking the weights
$\omega_{kl}(x) = (1-x)^{2+k} x^l (1+2x)$ with $(k,l)=\{(0,0), (1,0), (1,1), (1,2), (1,3)\}$  and $s_0=m_\tau^2$. 
With five moments, one can make a global fit of $\alpha_s(m_\tau^2)$, the gluon condensate, $\cO_6$ and $\cO_8$. To assess possible errors associated with neglected higher-order condensates, a second fit including $\cO_{10}$ has been performed and the variation on the fitted value of the strong coupling has been included as an additional uncertainty.
A quite precise value of $\alpha_{s}(m_{\tau}^{2})$ is obtained, in good agreement with Ref.~\cite{Davier:2013sfa}.
The extracted condensates have large relative errors exhibiting a very little sensitivity to power corrections. 
This has been further verified, taking away from the weights the factor $( 1+ 2x)$ which eliminates the highest-dimensional condensate contribution to every moment. This gives the fitted values shown in the second line of table~\ref{tab:summary}, which are in perfect agreement with the results of the previous fit (first line) and are even more precise. 

The doubly-pinched weights
$\omega^{(2,m)}(x) = (1-x)^2\sum_{k=0}^{m} (k+1)\, x^k  =
1-(m + 2)\, x^{m + 1} + (m + 1)\, x^{m + 2}$
are only sensitive to $\cO_{2 (m+2)}$ and $\cO_{2 (m+3)}$. A combined fit of five different $A^{(2,m)}$ moments ($1\le m\le 5$) gives the results shown in the third line of table~\ref{tab:summary}. First, a global fit with four free parameters, assuming $\cO_{12}=\cO_{14}=\cO_{16}=0$, has been done. To account for these missing power corrections, the fit has been repeated with the inclusion of $\cO_{12}$ and the variation in the fitted value of $\alpha_{s}(m_{\tau}^{2})$ has been taken as an additional uncertainty. The agreement with the results obtained in the previous fits is excellent.
Similar results (not included in the table) are obtained from a global fit to four $A^{(n,0)}$ ($0\le n\le 3$) moments based on the n-pinched weights
$\omega^{(n,0)}(x) = (1-x)^n$
which receive corrections from all condensates with $D\le 2(n+1)$, but are protected against duality violations for $n\not= 0$.

Neglecting all non-perturbative effects, one can determine $\alpha_{s}(m_{\tau}^{2})$ from a single moment. This interesting exercise has been also done in Ref.~\cite{Pich:2016bdg}, making 13 separate extractions of the strong coupling with six $A^{(2,m)}$ moments ($0\le m\le 5$), six $A^{(1,m)}$ moments ($0\le m\le 5$) based on the weights
$\omega^{(1,m)}(x) = 1-x^{m+1} = (1-x)\,\sum_{k=0}^{m} x^k$
which are only sensitive to $\cO_{2 (m+2)}$, and the moment $A^{(0,0)}$ where OPE corrections are absent but it is very exposed to duality-violation effects. In all cases, the resulting determinations of the strong coupling are in agreement with the values in table~\ref{tab:summary}, reflecting the minor numerical role of the neglected non-perturbative corrections.

\begin{figure}[t]
\centering
\includegraphics[width=0.46\textwidth]{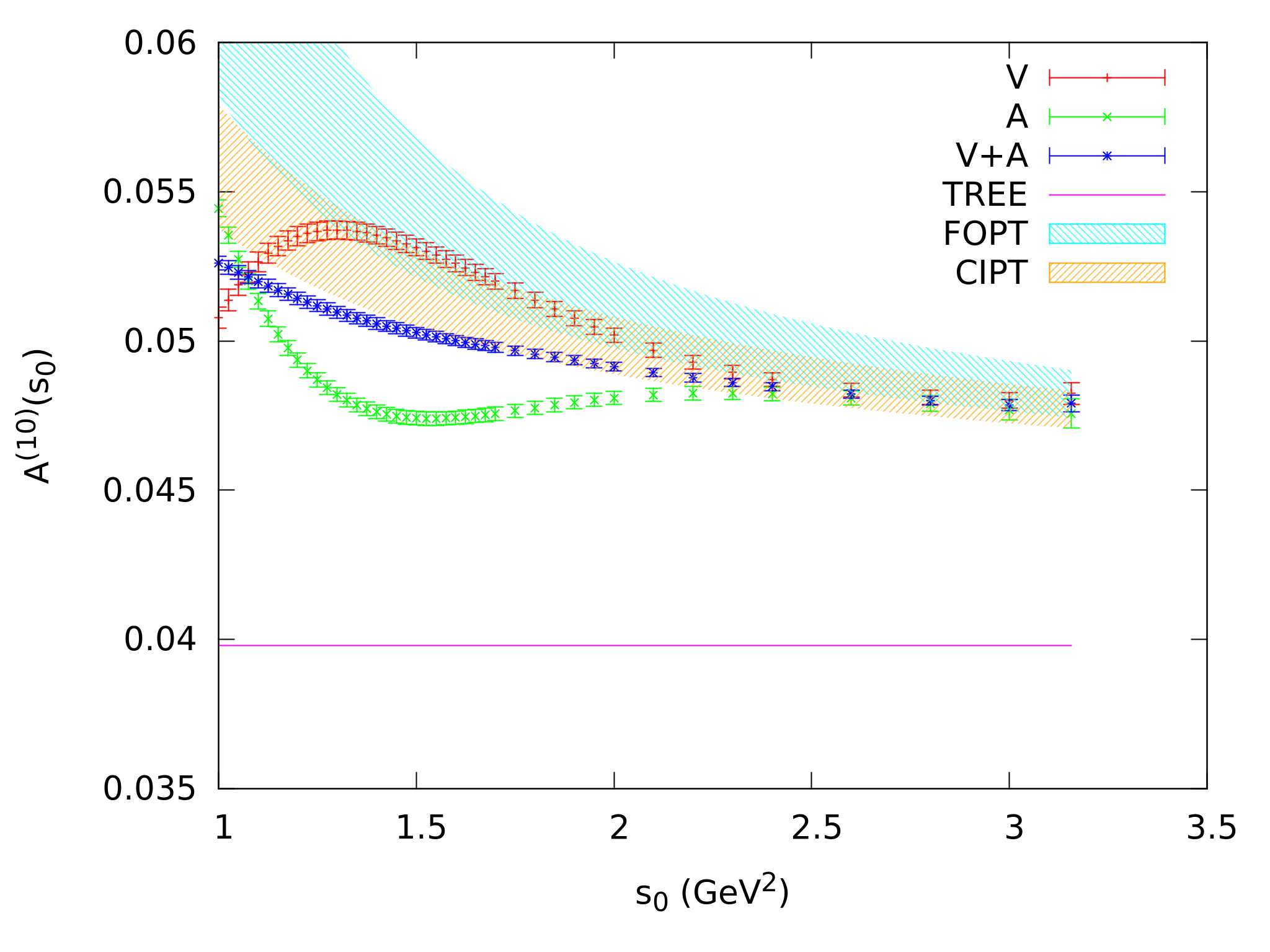} \hskip .5cm
\includegraphics[width=0.46\textwidth]{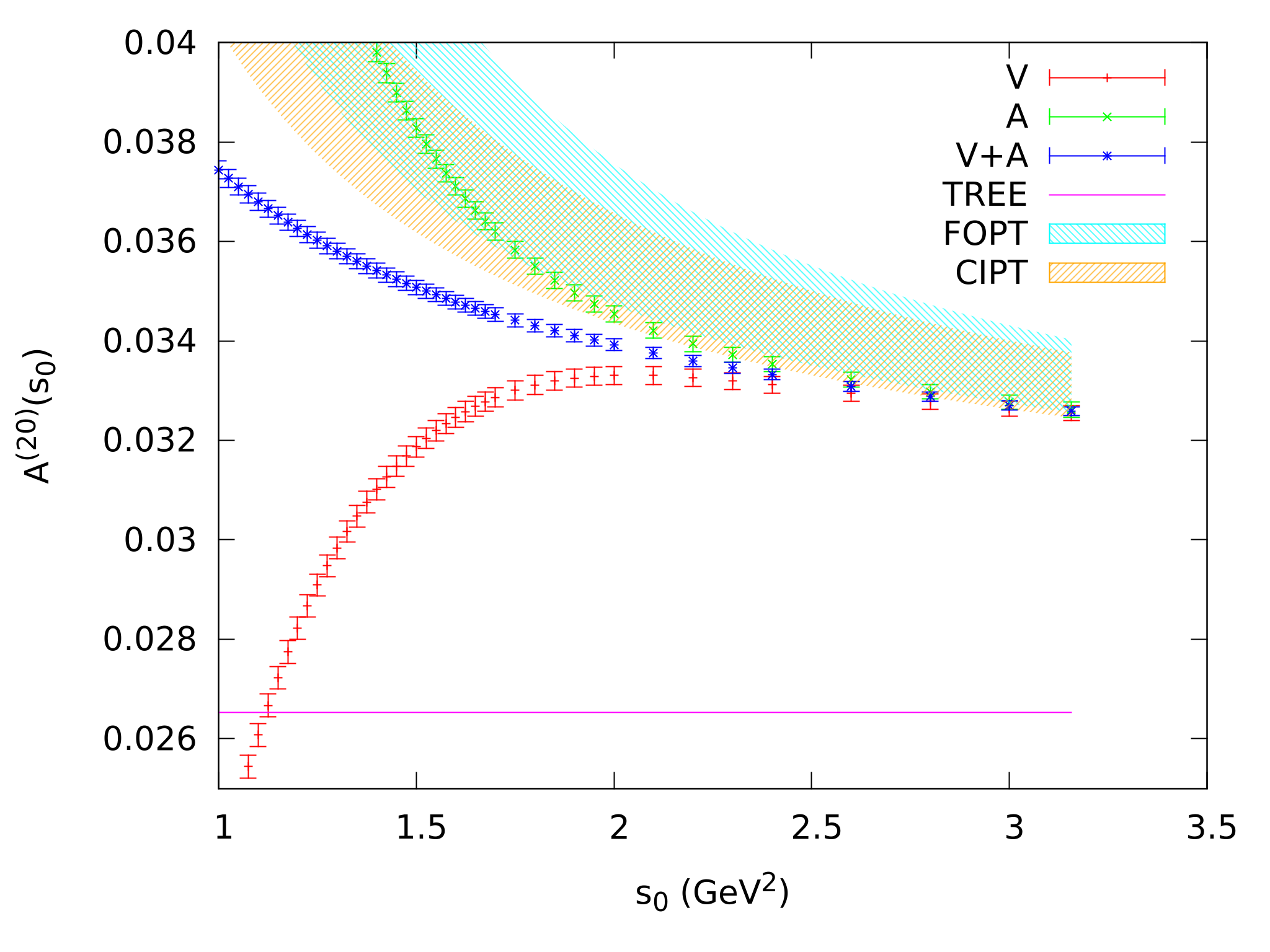}
\caption{Dependence on $s_0$ of the experimental moments $A^{(1,0)} (s_{0})$ (left) and $A^{(2,0)} (s_{0})$ (right), together with their purely CIPT and FOPT perturbative predictions for $\alpha_s^{(n_f=3)}(m_\tau^2) = 0.329 \,{}^{+\, 0.020}_{-\, 0.018}$.
Data points are shown for the $V$ (red), $A$ (green) and $\frac{1}{2}\, (V+A)$ (blue)
channels. The horizontal (pink) line indicates the free-parton result \protect\cite{Pich:2016bdg}.}
\label{pinchs}
\end{figure}

Non-perturbative contributions should manifest in a distinctive $s_0$ dependence.
Figure~\ref{pinchs} shows as function of $s_0$ the experimental moments $A^{(1,0)} (s_{0})$ and $A^{(2,0)} (s_{0})$, in the $V$, $A$ and $\frac{1}{2}\, (V+A)$ channels, together with their predicted values with
$\alpha_s^{(n_f=3)}(m_\tau^2) = 0.329 \,{}^{+\, 0.020}_{-\, 0.018}$, neglecting all non-perturbative contributions.
$A^{(1,0)} (s_{0})$, which can only get corrections from $\cO_4$, exhibits a surprisingly good agreement with its pure perturbative prediction. 
In spite of being only protected by a single pinch factor, the data points above $s_0\sim 2\;\mathrm{GeV}^2$ closely follow the central values predicted by CIPT. In that energy range non-perturbative contributions appear to be too small to become numerically visible within the much larger perturbative uncertainties covering the shades areas of the figure.
The splitting at lower values of $s_0$ of the $V$ and $A$ moments must be assigned to
duality violations, since their $D=4$ power corrections are approximately equal. However, these duality-violation effects clearly compensate in $V+A$, with an impressively flat distribution of the experimental data which does not deviate from the $1\sigma$ perturbative range even at $s_0 \sim 1\;\mathrm{GeV}$. A similar behaviour is observed for $A^{(0,0)} (s_{0})$, a moment without OPE corrections.
$A^{(2,0)} (s_{0})$ looks slightly more sensitive to non-perturbative contributions and seems to prefer a power correction with different signs for $V$ and $A$, which cancels to a good extend in $V+A$. This fits nicely with the expected $\cO_{6,V/A}$ contribution, although the merging of the $V$, $A$ and $V+A$  curves above $s_0\sim 2.2\;\mathrm{GeV}^2$ suggests a very tiny numerical effect from this source at high invariant masses.

Fitting the $s_0$ dependence of a single $A^{(2,m)} (s_{0})$ moment, one can determine the values of $\alpha_s(m_\tau^2)$, $\cO_{2(m+2)}$ and $\cO_{2(m+3)}$. 
The sensitivity to power corrections is very bad, as expected, but one finds an amazing stability in the extracted values of $\alpha_s(m_\tau^2)$. Including the information from the three lowest moments ($m=0,1,2$) and the nine energy bins above $s_0 = 2.0\;\mathrm{GeV}^2$, and adding as an additional uncertainty the small fluctuations observed when changing the number of fitted bins, one obtains the values of $\alpha_s(m_\tau^2)$ quoted in the fourth line of table~\ref{tab:summary}. 
Although they are much more sensitive to violations of quark-hadron duality 
(fitting the $s_0$ dependence of several consecutive bins, one is using information about the local structure of the spectral function), these results turn out to be in 
excellent agreement with the more solid determinations in the first three lines of the table.
The very flat shape of the $V+A$ hadronic distribution above $s_0 = 2.0\;\mathrm{GeV}^2$ implies small duality-violation effects in that region which, moreover, are very efficiently suppressed in the doubly-pinched moments $A^{(2,m)} (s_{0})$.

The marginal role of power corrections has been also corroborated, making independent 
$\alpha_s(m_\tau^2)$ determinations from seven $A^{(1,m)}(s_{0})$ ($0\le m\le 6$) and six $A^{(2,m)}(s_{0})$ ($0\le m\le 5$) moments, as function of $s_0$ and ignoring all non-perturbative effects.  In spite of the fact that these 13 moments get completely different OPE corrections, carrying a broad variety of inverse powers of $s_0$, all results exhibit a similar functional dependence on $s_0$. The small fluctuations among the different moments
stay in all cases well within the much larger perturbative uncertainties shown in figure~\ref{pinchs}.

Using weights of the type
$\omega_{a}^{(1,m)}(x) = (1- x^{m+1})\, \mathrm{e}^{-ax}$,
one suppresses potential violations of duality because the exponential factor nullifies the highest invariant-mass region, but paying the price that all condensates contribute to every moment. For $a=0$ one recovers the $A^{(1,m)}(s_0)$ moments, only affected by $\cO_{2 (m+2)}$, while for $a\gg 1$ the moments become independent of $m$. Thus, if one neglects all non-perturbative contributions, the OPE corrections should manifest in a larger instability under variations of $s_0$ than in the $a=0$ case. However, with $a\not= 0$ one gets even more stable results, and the different moments converge very soon when $a$ increases, indicating again that power corrections are not very relevant. From the analysis of seven $V+A$ moments ($m=0,\cdots,6$), accepting for each moment all values of $\alpha_{s}(m_{\tau}^{2})$ in the Borel-stable region, and adding as additional theoretical uncertainties the differences among moments and the variations in the region $s_0\in [2,2.8]\;\mathrm{GeV}^2$), one gets the determination of $\alpha_{s}(m_{\tau}^{2})$ shown in the fifth line of table~\ref{tab:summary}.

\subsection{Violations of quark-hadron duality}

The small differences between the true values of the moments $A^\omega_J(s_0)$ and their OPE approximations are known as (global) duality violations. Using analyticity, they can be fomally expressed as  \cite{Cata:2008ye,Chibisov:1996wf,GonzalezAlonso:2010rn,Rodriguez-Sanchez:2016jvw}
\be\label{eq:DVcorr}
\Delta A^{\omega, \mathrm{DV}}_{J}(s_{0})\;\equiv\;
\frac{i}{2}\;\oint_{|s|=s_{0}}
\frac{ds}{s_{0}}\;\omega(s)\,\left\{\Pi_{ud,J}^{(0+1)}(s) - \Pi_{ud,J}^{(0+1)}(s)^{\mathrm{OPE}}
\right\}
\; =\; -\pi\;\int^{\infty}_{s_{0}} \frac{ds}{s_{0}}\;\omega(s)\; 
\Delta\rho^{\mathrm{DV}}_{J}(s)
\, ,
\ee
with $\Delta\rho^{\mathrm{DV}}_{V/A}(s)$ the differences between the physical spectral functions and their OPE estimates which, unfortunately, are unknown beyond the experimentally accessed region. Owing to asymptotic freedom, the violations of duality should decrease very fast as $s_0$ increases. In practice, they are minimized by taking ``pinched'' weight functions which vanish at $s=s_0$ and suppress the contributions from the region near the real axis where the OPE is not valid \cite{Braaten:1991qm,LeDiberder:1992zhd}. The many tests discussed before clearly indicate that these effects are negligible in the extraction of $\alpha_s(m_\tau^2)$ from the $V+A$ distribution.

Instead of using clean moments where duality violations are suppressed, some works focus on observables more sensitive to these uncontrollable effects \cite{Boito:2014sta}, 
modelling them with an ansatz for $\Delta\rho^{\mathrm{DV}}_{J}(s)$ which is fitted to the measured spectral functions. Since the OPE is not valid on the physical cut, one loses theoretical control and gets at best an effective model description with unclear relation with QCD.
Let us consider the slightly generalized ansatz (in GeV units)
\be\label{eq:DVparam}
\Delta\rho^{\mathrm{DV}}_{J}(s)\; =\; s^{\lambda_{J}}\; e^{-(\delta_{J}+\gamma_{J}s)}\;\sin{(\alpha_{J}+\beta_{J}s)}\, ,
\qquad\qquad
s> \hat s_0\, ,
\ee
which for $\lambda_{J} =0$ coincides with the model assumed in Ref.~\cite{Boito:2014sta}.
The combination of a dumping exponential with an oscillatory function is expected to describe the fall-off of duality violations at very high energies, but this functional form is completely ad-hoc and difficult to justify at low energies.

\begin{figure}[t]
\centering
\begin{minipage}[c]{0.45\textwidth}\mbox{}\hskip -.4cm
\includegraphics[width=\textwidth]{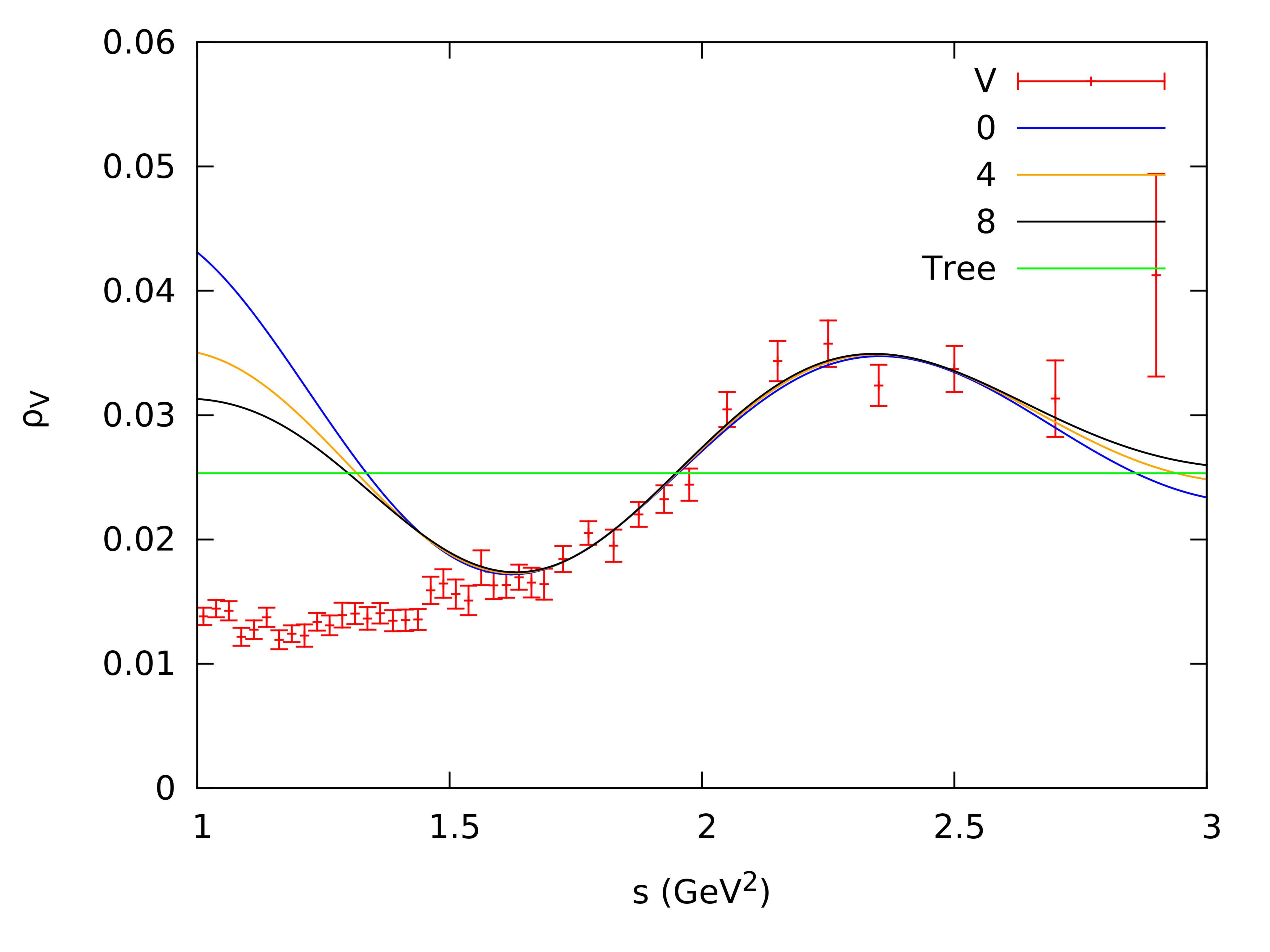}
\end{minipage}
\hskip .5cm
\begin{minipage}[c]{0.45\textwidth}\centering
\begin{tabular}{|c|c|c|c|c|}\hline &&&&\\[-10pt]
$\lambda_V$  & $\alpha_{s}(m_{\tau}^{2})$ & $\delta_V$      & $\gamma_V$     
& p-value
\\[1.2pt] \hline &&&&\\[-10pt]
0  & $0.298 \; (10)$           & $3.6 \; (5)$ & $0.6 \; (3)$ 
& 5.3 \%
\\ 
1  & $0.300 \; (12)$          & $3.3 \; (5)$ & $1.1 \; (3)$ 
& 5.7 \%
\\ 
2  & $0.302 \; (11)$          & $2.9 \; (5)$ & $1.6 \; (3)$ 
& 6.0 \%
\\ 
4  & $0.306 \; (13)$          & $2.3 \; (5)$ & $2.6 \; (3)$ 
& 6.6 \%
\\
8  & $0.314 \; (15)$          & $1.0 \; (5)$ & $4.6 \; (3)$ 
& 7.7 \%
\\ \hline
\end{tabular}
\end{minipage}
\caption{Vector spectral function $\rho_V^{\protect\phantom{()}}(s)$, fitted
with the ansatz (\ref{eq:DVparam}) for different values of $\lambda_V$, compared with the data points. The right table shows a representative subset of the fitted parameters with FOPT \cite{Pich:2016bdg}.}
\label{fig:spectral-n}
\end{figure}

Since there are far too many parameters to be fitted to a highly-correlated data set, Ref.~\cite{Boito:2014sta} concentrates in the moment $A^{(0,0)}_{V}(s_{0})$ which is very exposed to violations of duality ($\omega(x)=1$) and does not receive OPE corrections (owing to the tail of the $a_1$ resonance, the axial channel is not very useful).
The  model parameters and $\alpha_s$ are determined fitting the $s_0$ dependence for $s_0\ge \hat s_0 = 1.55~\mathrm{GeV}^2$. This choice has the largest, but still too small, p-value and gives the smallest $\alpha_s$. However, the p-value falls dramatically when one moves from this point, becoming worse at higher $\hat s_0$ values where the model should work better. The extracted value of $\alpha_s$ is very unstable under small modifications of the fit procedure and the fitted ansatz strongly deviates from the data as soon as one moves from the fitted region. This is illustrated in figure~\ref{fig:spectral-n} which shows the results of this exercise with FOPT, for different values of the power $\lambda_{V}$ and $\hat s_0 = 1.55~\mathrm{GeV}^2$.
The actual uncertainties are much larger than the quoted fit errors; varying $\hat s_0$ in the range $[1.15 , 1.75]~\mathrm{GeV}^2$, with $\lambda_V=0$, induces $3\sigma$ fluctuations of $\alpha_s(m_\tau^2)$.

All models reproduce well $\rho_V^{\protect\phantom{()}}(s)$ in the fitted region ($s\ge 1.55\;\mathrm{GeV}^2$), but they fail badly below it. The choice $\lambda_V=0$ assumed in Ref.~\cite{Boito:2014sta} is clearly the worse one. Increasing the power $\lambda_V$, the ansatz  slightly approaches the data below the fitted range, while the exponential parameters $\delta_V$ and $\gamma_V$ adapt themselves to compensate the growing at high values of $s$ with the net result of a smaller duality-violation correction. The statistical quality of the fit improves also with growing values of $\lambda_V$, while 
$\alpha_s(m_\tau^2)$ increases approaching the more solid FOPT values in table~\ref{tab:summary}. 
The strong correlation of the fitted $\alpha_s(m_\tau^2)$ with the assumed model should not be a surprise because one is just fitting models to data without any strong theoretical guidance (the OPE is no longer valid), and $\alpha_s$ has been converted into one more model parameter. In spite of all caveats, one gets still quite reasonable values of the strong coupling, but they are model dependent and, thus, unreliable.

\subsection{Updated determination of $\boldsymbol{\alpha_s(m_\tau^2)}$}

The results shown in table~\ref{tab:summary} are based on solid theoretical principles
(the $s_0$-dependence extraction assumes, however, local duality) and exhibit a good stability under small variations of the fit procedures. The overall agreement among determinations extracted under very different assumptions shows their reliability and even indicates that the uncertainties are probably too conservative.
Averaging the five determinations, but keeping the smaller uncertainties to account for the large correlations, one finds
\be
\alpha_{s}^{(n_f=3)}(m_\tau^2)^{\mathrm{CIPT}} \; =\; 0.335 \pm 0.013\, ,
\qquad\qquad\quad
\alpha_{s}^{(n_f=3)}(m_\tau^2)^{\mathrm{FOPT}} \; =\; 0.320 \pm 0.012\, .
\label{FinalValues}
\ee
The same results are obtained irrespective or whether one includes or not in the average the determination from the $s_0$ dependence of the moments. Averaging the CIPT and FOPT ``averages'' in table~\ref{tab:summary}, one finally gets
\be
\alpha_{s}^{(n_f=3)}(m_\tau^2) \; =\; 0.328 \pm 0.013 \, .
\ee
These results nicely agree with the value of the strong coupling extracted from $R_\tau$ \cite{Pich:2013lsa}.

After evolution up to the scale $M_Z$, the strong coupling decreases to
\be
\alpha_{s}^{(n_f=5)}(M_Z^{2})\; =\; 0.1197\pm 0.0015 \, ,
\ee
in excellent agreement with the direct measurement at the $Z$ peak in Eq.~(\ref{eq:alpha_Z}).
The comparison of these two determinations, graphically shown in the right panel of figure~\ref{fig:running}, provides a beautiful test of the predicted QCD running; {\it i.e.}, a very significant experimental verification of asymptotic freedom:
\be
\left.\alpha_{s}^{(n_f=5)}(M_Z^{2})\right|_\tau - \left.\alpha_{s}^{(n_f=5)}(M_Z^{2})\right|_Z
\; =\; 0.0001\pm 0.0015_\tau\pm 0.0030_Z\, .
\ee

Improvements on the determination of $\alpha_{s}(m_\tau^2)$ from $\tau$ decay data would require high-precision measurements of the spectral functions, specially in the higher kinematically-allowed energy bins. Both higher statistics and a good control of experimental systematics are needed, which could be possible at the forthcoming Belle-II experiment. On the theoretical side, one needs an improved understanding of higher-order perturbative corrections.

\section{World average value of $\boldsymbol{\alpha_s}$}
\label{sec:worldaverage}

\begin{figure}[t]
\centering
\includegraphics[height=4.25cm,clip]{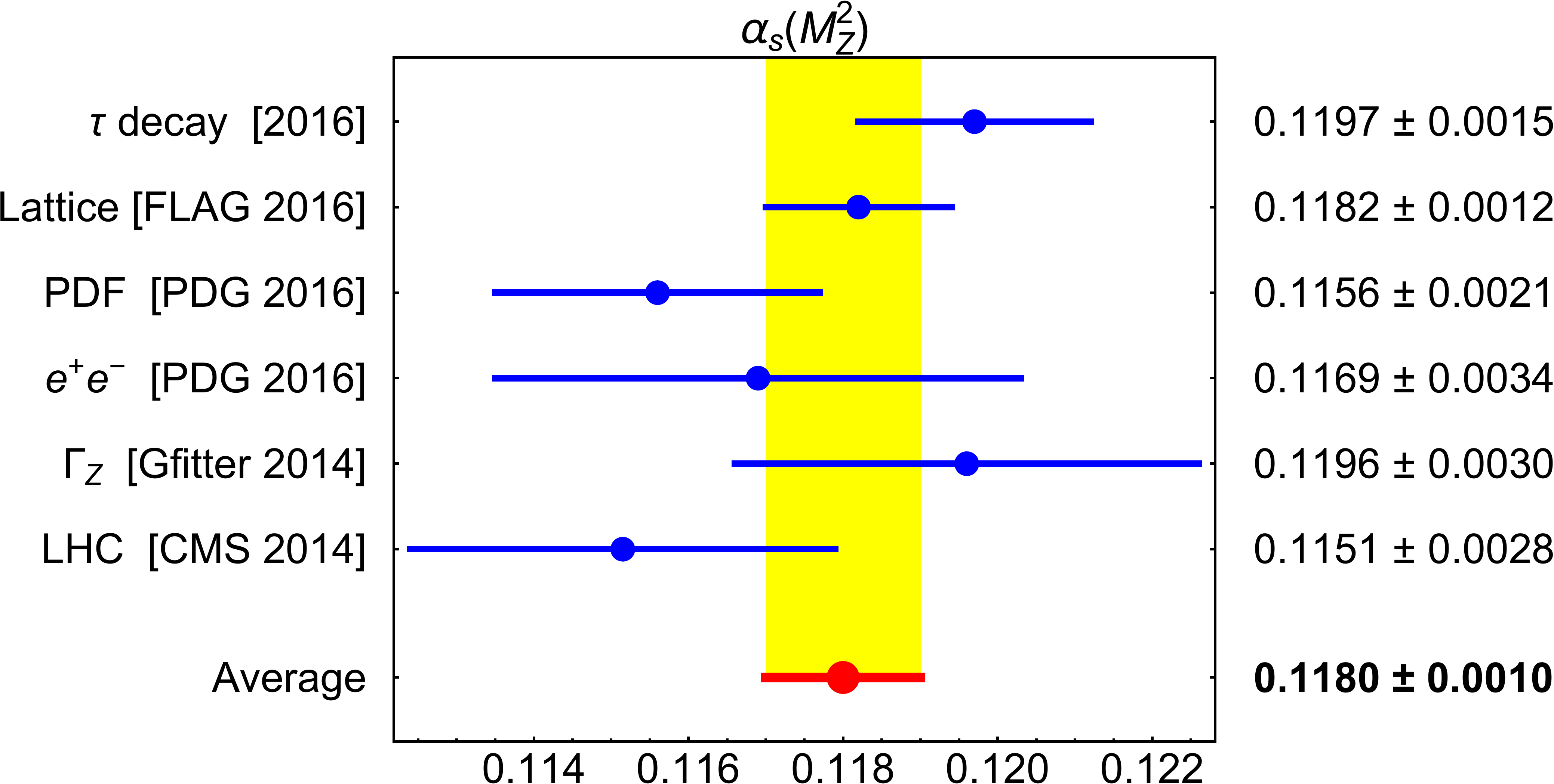}
\caption{Summary of the most precise determinations of $\alpha_s(M_Z^2)$.} 
\label{fig:alpha-wav}
\end{figure}

Figure~\ref{fig:alpha-wav} compares the N${}^3$LO determinations of $\alpha_s(M_Z^2)$ from $Z$ and $\tau$ decays with other precise measurements of the strong coupling. Following the PDG criteria \cite{BDS:16}, only those determinations which are at least of NNLO are taken into account. This includes several event-shape analyses in hadronic final states of $e^+e^-$ annihilations, and studies of parton distribution functions from deep inelastic scattering and hadron collider data. The numbers quoted in the figure correspond in both cases to the recent PDG compilation \cite{BDS:16}.

The PDG includes also in the average the CMS determination from the $t\bar t$ production cross section at $\sqrt{s}=7$~TeV, $\alpha_s(M_Z^2)= 0.1151\,{}^{+\, 0.0028}_{-\, 0.0027}$ \cite{Chatrchyan:2013haa}, which requires as input a value of the top quark mass
(either $\alpha_s$ or $m_t$ are fitted to the data, but not both). 
Although there are more recent measurements of this cross section from ATLAS and CMS, at $\sqrt{s} = 7,\, 8$ and 13 TeV, none of them quotes further determinations of  $\alpha_s$. Applying the same procedure, these measurements would imply larger values of $\alpha_s(M_Z^2)$ than the one of Ref.~\cite{Chatrchyan:2013haa}, which is nevertheless included in the average.

The most precise value of $\alpha_s(M_Z^2)$ is obtained from lattice simulations, with a growing number of groups measuring (non-perturbatively) various short-distance quantities, through numerical evaluations of the QCD functional integral, and comparing the results with the corresponding perturbative expansions in powers of the strong coupling. The present situation has been recently summarized by the FLAG working group \cite{Aoki:2016frl} which quotes the lattice world-average shown in figure~\ref{fig:alpha-wav}.

The different determinations in figure~\ref{fig:alpha-wav} are in good agreement, within their quoted errors. From these results, one obtains the final world average value
\be
\alpha_s(M_Z^2)\; =\; 0.1180 \pm 0.0010\, .
\ee
This number is very close to the 2016 PDG average ($0.1181\pm 0.0011$), which does not yet include the most recent $\tau$ decay and lattice results.
The central value has been directly obtained as the weighted average of the six input determinations, while the error has been enlarged applying the PDG prescription, {\it i.e.}, adjusting all individual uncertainties by a common factor such that $\chi^2/\mathrm{dof}$ equals unity.
Removing the CMS determination would slightly increase the central value, giving as average
$\alpha_s(M_Z^2) = 0.1183 \pm 0.0011$. The overall uncertainty is largely determined by the precise lattice result.

\begin{acknowledgement}
I would like to thank Antonio Rodríguez Sánchez for a very enjoyable collaboration.
This work has been supported in part by the Spanish Government and ERDF funds from
the EU Commission [Grant FPA2014-53631-C2-1-P], by the Spanish
Centro de Excelencia Severo Ochoa Programme [Grant SEV-2014-0398] and by the Generalitat Valenciana [Grant PrometeoII/2013/007].

\end{acknowledgement}

\end{document}